\documentclass[a4paper,twoside,12pt]{article}
\input{obsjkt.sty}

\newcommand{\Msun}{\ensuremath{~{\rm M}_\odot}}                   
\newcommand{\Rsun}{\ensuremath{~{\rm R}_\odot}}                   
\newcommand{\rhosun}{\ensuremath{~\rho_\odot}}                    
\newcommand{\Teff}{\ensuremath{T_{\rm eff}}}                      
\newcommand{\EBV}{\ensuremath{E(B\!-\!V)}}                        
\newcommand{\Grp}{\ensuremath{G_{\rm RP}}}                        
\newcommand{\degr}{\ensuremath{^\circ}}                           
\newcommand{\Vsini}{\ensuremath{V \sin i}}                        
\renewcommand{\kms}{~km~s$^{-1}$}                                 
\newcommand{\chisq}{\ensuremath{\chi^{\,2}}}                      
\newcommand{\chir}{\ensuremath{\chi_\nu^{\,2}}}                   
\newcommand{\hip}{\textit{Hipparcos}}                             
\newcommand{\gaia}{\textit{Gaia}}                                 
\newcommand{\targ}{DV~Boo}
\newcommand{\targfull}{DV~Bo\"otes}

\newcommand{\Msunnom}{\hbox{$\mathcal{M}^{\rm N}_\odot$}}
\newcommand{\Rsunnom}{\hbox{$\mathcal{R}^{\rm N}_\odot$}}
\newcommand{\Lsunnom}{\hbox{$\mathcal{L}^{\rm N}_\odot$}}

\usepackage{rotating}

\begin{document} 

\OBSheader{Rediscussion of eclipsing binaries: \targ}{J.\ Southworth}{2026 February}

\OBStitle{Rediscussion of eclipsing binaries. Paper XXVIII. \\ The metallic-lined system DV~Bo\"otes}

\OBSauth{John Southworth}

\OBSinstone{Astrophysics Group, Keele University, Staffordshire, ST5 5BG, UK}


\OBSabstract{\targ\ is a detached eclipsing binary containing a metallic-lined A-star and a chemically normal late-F star, in an orbit with a period of 3.783~d and a possible slight eccentricity. We use a light curve from the Transiting Exoplanet Survey Satellite (TESS) and published spectroscopic results to determine the physical properties of the system to high precision. We find masses of $1.617 \pm 0.003$\Msun\ and $1.207 \pm 0.004$\Msun, and radii of $1.948 \pm 0.008$\Rsun\ and $1.195 \pm 0.022$\Rsun. The precision of the radius measurements is limited by the shallow partial eclipses and the unavailability of a spectroscopic light ratio due to the chemical peculiarity of the primary star. We measure a distance to the system of $125.0 \pm 1.5$~pc, in good agreement with the \gaia\ DR3 parallax, and an age of 1.3~Gyr. A comparison with theoretical models suggests the system has a modestly sub-solar metallicity, in conflict with the slightly super-solar photospheric abundances of the secondary star.}


\section*{Introduction}

Detached eclipsing binaries (dEBs) are a valuable source of direct measurements of the basic physical properties of normal stars \cite{Andersen91aarv,Torres++10aarv,Me15aspc}. From light and radial velocity (RV) curves it is possible to measure their masses and radii directly using geometry and celestial mechanics, and without reliance on theoretical stellar models. The current series of papers \cite{Me20obs} is dedicated to using new space-based \cite{Me21univ} light curves of a significant number of dEBs to improve measurements of their physical properties.

In this work we present a study of \targfull\ (Table~\ref{tab:info}), which consists of a metallic-lined (Am) primary component (hereafter star~A) and a late-F secondary component (star~B) which appears to be chemically normal. Such objects are well represented in the list of well-studied dEBs \cite{Me15aspc} because a high fraction of Am stars are in short-period binaries \cite{Abt61apjs,Abt65apjs,CarquillatPrieur07mn} and because A-stars in dEBs are comparatively easy to study due to their quiet photospheres (no magnetic activity and usually no pulsations) and modest rotational velocities (allowing precise RV measurements).


\section*{\targfull}

\begin{table}[t]
\caption{\em Basic information on \targfull. 
The $BV$ magnitudes are each the mean of 92 individual measurements \cite{Hog+00aa} distributed approximately randomly in orbital phase. 
The $JHK_s$ magnitudes are from 2MASS \cite{Cutri+03book} and were obtained at an orbital phase of 0.13. \label{tab:info}}
\centering
\begin{tabular}{lll}
{\em Property}                            & {\em Value}                 & {\em Reference}                      \\[3pt]
Right ascension (J2000)                   & 14 22 49.698                & \citenum{Gaia23aa}                   \\
Declination (J2000)                       & +14 56 20.14                & \citenum{Gaia23aa}                   \\
Henry Draper designation                  & HD 126931                   & \citenum{CannonPickering20anhar}     \\
\textit{Hipparcos} designation            & HIP 70287                   & \citenum{ESA97}                      \\
\textit{Tycho} designation                & TYC 915-464-1               & \citenum{Hog+00aa}                   \\
\textit{Gaia} DR3 designation             & 1228635253980613504         & \citenum{Gaia21aa}                   \\
\textit{Gaia} DR3 parallax (mas)          & $7.9495 \pm 0.0274$         & \citenum{Gaia21aa}                   \\          
TESS\ Input Catalog designation           & TIC 450349567               & \citenum{Stassun+19aj}               \\
$B$ magnitude                             & $7.945 \pm 0.009$           & \citenum{Hog+00aa}                   \\          
$V$ magnitude                             & $7.578 \pm 0.009$           & \citenum{Hog+00aa}                   \\          
$J$ magnitude                             & $6.836 \pm 0.027$           & \citenum{Cutri+03book}               \\
$H$ magnitude                             & $6.735 \pm 0.029$           & \citenum{Cutri+03book}               \\
$K_s$ magnitude                           & $6.704 \pm 0.020$           & \citenum{Cutri+03book}               \\
Spectral type                             & kA4hF1mF3(V) + F6/7V        & \citenum{McGahee+20aj}, \citenum{Carquillat+04mn}               \\[3pt]       
\end{tabular}
\end{table}



\targ\ was found to be eclipsing using photometry from the \hip\ satellite \cite{ESA97}, and was given its variable-star designation by Kazarovets et al.\ \cite{Kazarovets+99ibvs}. Bidelman \cite{Bidelman88pasp} specified it as an Am star; Grenier et al.\ \cite{Grenier+99aas} classified it as A3mA7F5, and McGahee et al.\ \cite{McGahee+20aj} updated this to kA4hF1mF3(V) following the standard approach of giving spectral classes for chemically-peculiar stars based on their Ca~I K-line, hydrogen lines, and metal lines.

Carquillat et al.\ \cite{Carquillat+04mn} obtained the first spectroscopic orbit of \targ, based on data from three spectrographs (\textit{\'Elodie} plus two \textit{Coravel} instruments) and comprising 48 radial velocity measurements (RVs) for star~A and 10 RVs for star~B. They found a spectroscopic light ratio of $0.41 \pm 0.05$ from the ratio of the cross-correlation dips in their \textit{\'Elodie} spectra, which corresponds to the ratio of the spectral line strengths of the two components. This is not the same as a continuum light ratio due to the change in intrinsic line strength with temperature as well as the effect of the chemical peculiarity of star~A. Carquillat et al.\ determined effective temperature (\Teff) values of $7370 \pm 80$~K and $6410 \pm 80$~K, and a projected rotational velocity for star~A of $\Vsini = 24.4 \pm 2.4$\kms. These authors also fitted the \hip\ light curve of the system to determine masses and approximate radii for the two components.

       
Kahraman Ali{\c{c}}avu{\c{s}} \& Ali{\c{c}}avu{\c{s}} \cite{KahramanAlicavus20raa} presented an updated analysis of \targ\ based on the spectra from \textit{\'Elodie}, additional archival spectra from the FEROS and HARPS \'echelle spectrographs, and three light curves from small survey telescopes. They obtained mass and radius measurements to 0.25\% and 2.5\%, respectively, and \Vsini\ values of $26 \pm 2$ and $17 \pm 3$\kms. A detailed abundance analysis confirmed that star~A is a typical Am star with underabundances of Ca and Sc and overabundances of iron-peak elements.


Catanzaro et al.\ \cite{Catanzaro+24aa} provided the most recent analysis of \targ, using a further 16 spectra from the CAOS \'echelle spectrograph at Catania Astrophysical Observatory. These authors were the first to have access to a high-quality light curve of the system, from the Transiting Exoplanet Survey Satellite \cite{Ricker+15jatis} (TESS), which was modelled together with the RVs to determine the properties of the component stars. They (re)confirmed that star~A is an Am star, found star~B to have a normal photospheric chemical composition, and ruled out the existence of $\delta$\,Scuti pulsations in the system.


\section*{Photometric observations}


\begin{figure}[t] \centering \includegraphics[width=\textwidth]{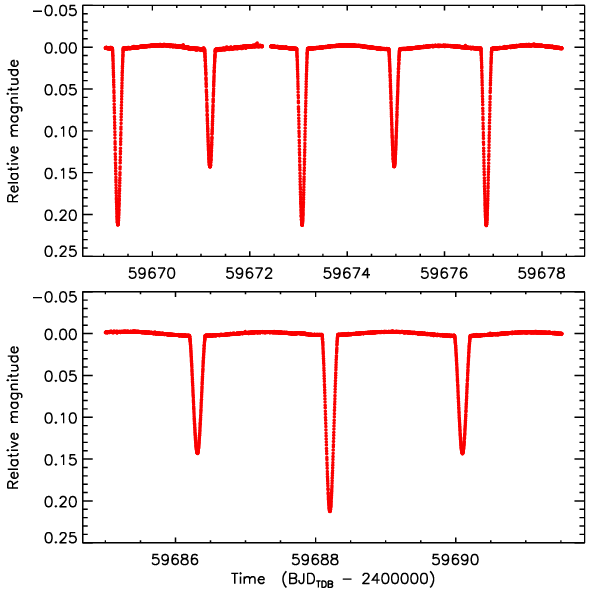} \\
\caption{\label{fig:time} TESS sector 50 photometry of \targ, including only the data analysed in the current 
work. The flux measurements have been converted to magnitude units and the median subtracted.} \end{figure}

\targ\ has so far been observed by TESS in just one sector, 50, at a cadence of 120~s. We downloaded the SPOC (Science Processing Center \cite{Jenkins+16spie}) light curve from the NASA Mikulski Archive for Space Telescopes (MAST\footnote{\texttt{https://mast.stsci.edu/portal/Mashup/Clients/Mast/Portal.html}}) using the {\sc lightkurve} package \cite{Lightkurve18}. A significant portion of the light curve is subject to quality flags, and specifying the ``hard'' option returns a total of 12,948 datapoints. These were converted into differential magnitude and the median magnitude was subtracted for convenience. Isolated portions of the light curve were then removed to leave two stretches of data containing five and three eclipses respectively, and totalling 11,273 datapoints. Fig.~\ref{fig:time} shows the light curve after this initial processing.

We queried the \gaia\ DR3 database\footnote{\texttt{https://vizier.cds.unistra.fr/viz-bin/VizieR-3?-source=I/355/gaiadr3}} for all sources within 2~arcmin of \targ. Only 13 more objects were found, the brightest of which is fainter by $\Grp = 9.3$~mag, so we do not expect a significant amount of contaminating light in the TESS light curves.

\section*{Light curve analysis}

\begin{table} 
\begin{center}
\caption{\em Times of mid-eclipse for \targ\ and their residuals versus the fitted ephemeris. 
The ephemeris zeropoint was chosen to be during the TESS observations. \label{tab:tmin}}
\setlength{\tabcolsep}{10pt}
\begin{tabular}{rllrl}
{\em Orbital} & {\em Eclipse time}  & {\em Uncertainty} & {\em Residual} & {\em Reference} \\
{\em cycle}   & {\em (BJD$_{TDB}$)} & {\em (d)}         & {\em (d)}      &                 \\[3pt]
$-3075.0$ & 2448045.254   & 0.005   & $-0.00107$ & \citenum{OteroDubovsky04ibvs} \\
$-3124.0$ & 2447859.9071  & 0.0005  & $ 0.00112$ & \citenum{Carquillat+04mn}     \\
$-1839.0$ & 2452720.5880  & 0.0010  & $-0.00343$ & \citenum{Porowski05ibvs}      \\
$-1153.0$ & 2455315.4755  & 0.0036  & $-0.00325$ & \citenum{Brat+11oejv}         \\
$-1066.0$ & 2455644.56349 & 0.00098 & $-0.00447$ & \citenum{Zasche+11ibvs}       \\
$ -477.0$ & 2457872.5360  & 0.0070  & $-0.00374$ & \citenum{Pasche18oejv}        \\
$ -476.5$ & 2457874.4289  & 0.0025  & $-0.00228$ & \citenum{Pagel18ibvs}         \\
$ -182.0$ & 2458988.4224  & 0.0028  & $ 0.00546$ & \citenum{Pagel21bavj}         \\
$    0.0$ & 2459676.856474& 0.000006&            & ~This work$^*$                 \\
$  103.0$ & 2460066.4670  & 0.0080  & $-0.00080$ & \citenum{Paschke23bav}        \\
\end{tabular}
\end{center}     
{\em $^*$ The eclipse time for the TESS observations was obtained from all data from sector 50. It was 
not included in the final {\sc jktebop} analysis to avoid double-use of data, but is given for reference.}\\
\end{table}


The components of \targ\ are well-separated and suitable for analysis with the {\sc jktebop}\footnote{\texttt{http://www.astro.keele.ac.uk/jkt/codes/jktebop.html}} code \cite{Me++04mn2,Me13aa}, for which we used version 44. We fitted for the following parameters: the fractional radii of the stars ($r_{\rm A}$ and $r_{\rm B}$) taken as the sum ($r_{\rm A}+r_{\rm B}$) and ratio ($k = r_{\rm B}/r_{\rm A}$), the central surface brightness ratio ($J$), third light ($L_3$), orbital inclination ($i$), orbital period ($P$), and a reference time of primary minimum ($T_0$). Limb darkening (LD) was accounted for using the power-2 law \cite{Hestroffer97aa,Maxted18aa,Me23obs2}, the linear coefficients ($c$) were fitted, and the non-linear coefficients ($\alpha$) were fixed at theoretical values \cite{ClaretSouthworth22aa,ClaretSouthworth23aa}. The measurement errors were scaled to force a reduced $\chi^2$ of $\chir = 1.0$. We additionally fitted for the coefficients of two first-order polynomials, one for each part of the light curve, to account for any slow brightness trends. 

\begin{figure}[t] \centering \includegraphics[width=\textwidth]{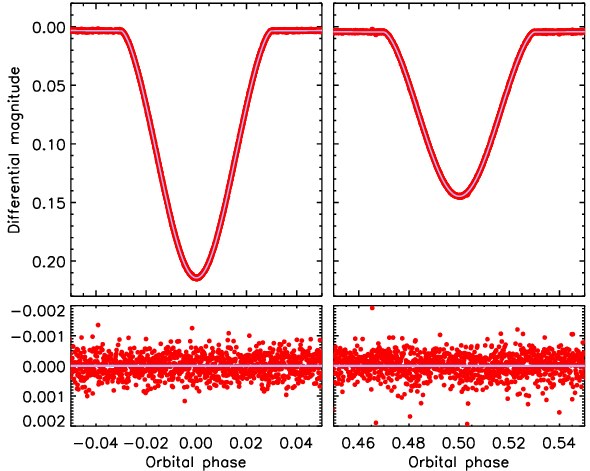} \\
\caption{\label{fig:phase} {\sc jktebop} best fit to the light curves of \targ\ from 
TESS sector 50 for the primary eclipse (left panels) and secondary eclipse (right panels). 
The data are shown as filled red circles and the best fit as a light blue solid line. 
The residuals are shown on an enlarged scale in the lower panels.} \end{figure}

We initially assumed a circular orbit, in line with previous analyses, but found that a better fit ($\chir = 11,366$ versus 11,473) could be obtained with a small amount of $e\cos\omega$, where $e$ is the eccentricity and $\omega$ is the argument of periastron. We therefore added both $e\cos\omega$ and $e\sin\omega$ to the list of fitted parameters. The relatively shallow partial eclipses (0.22 and 0.15 mag) and non-zero orbital eccentricity made it likely that the results of the light curve could be imprecise due to correlations between parameters. We therefore added some of the existing RVs to our analysis to provide more constraints on the shape and orientation of the orbit. After some experimentation we included the \textit{\'Elodie} RVs from ref.\ \citenum{Carquillat+04mn} and the CAOS RVs from ref.\ \citenum{Catanzaro+24aa}. For the latter we rejected the RVs from a spectrum taken at phase 0.980 due to blending effects. The fitted parameters were augmented with the velocity amplitudes and systemic velocities for each star.

\begin{figure}[t] \centering \includegraphics[width=\textwidth]{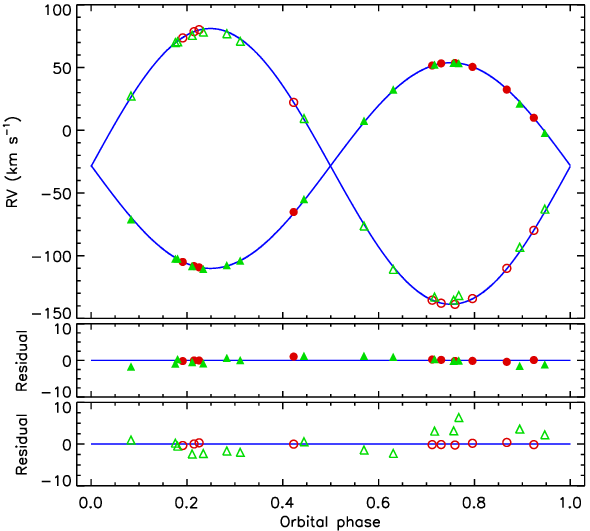} \\
\caption{\label{fig:rv} RVs of \targ\ compared to the best fit from the {\sc jktebop} 
analysis (solid blue lines). The RVs for star~A are shown with filled symbols, and for 
star~B with open symbols. The residuals are given in the lower panels separately for the 
two components. RVs from \textit{\'Elodie} \cite{Carquillat+04mn} are shown with red 
circles, and those from CAOS \cite{Catanzaro+24aa} with green triangles.} \end{figure}

Simultaneous analysis of the TESS data, obtained in April 2022, and RVs taken in the years 2001--2002 and 2014--2022 is helped by having a precise orbital ephemeris. We therefore included in our {\sc jktebop} fit nine times of minimum taken from the literature \cite{Kreiner++01book} (Table~\ref{tab:tmin}). Our fit therefore included the TESS light curve, RVs for both stars, and times of minimum covering 32.4 years. The fit to the light curve is shown in Fig.~\ref{fig:phase}, and to the RVs in Fig.~\ref{fig:rv}. Its parameters are given in Table~\ref{tab:jktebop}.

\begin{table} \centering
\caption{\em \label{tab:jktebop} Photometric parameters of \targ\ measured using {\sc jktebop} from 
TESS photometry and \textit{\'Elodie} and CAOS RVs. The uncertainties are $1\sigma$ errorbars.}
\begin{tabular}{lcc}
{\em Parameter}                           &              {\em Value}            \\[3pt]
{\it Fitted parameters:} \\
Orbital period (d)                        & $      3.7826346   \pm  0.0000006 $ \\
Time of primary eclipse (BJD$_{\rm TDB}$) & $ 2459676.856434   \pm  0.000012  $ \\
Orbital inclination (\degr)               & $      83.53       \pm  0.17      $ \\
Sum of the fractional radii               & $       0.2177     \pm  0.0015    $ \\
Ratio of the radii                        & $       0.613      \pm  0.011     $ \\
Central surface brightness ratio          & $       0.706      \pm  0.036     $ \\
Third light                               & $       0.025      \pm  0.019     $ \\
$e\cos\omega$                             & $       0.000051   \pm  0.000012  $ \\
$e\sin\omega$                             & $       0.0037     \pm  0.0049    $ \\
LD coefficient $c_{\rm A}$                & $       0.600      \pm  0.064     $ \\
LD coefficient $c_{\rm B}$                & $       0.680      \pm  0.049     $ \\
LD coefficient $\alpha_{\rm A}$           &              0.4030 (fixed)         \\
LD coefficient $\alpha_{\rm B}$           &              0.4984 (fixed)         \\
Velocity amplitude for star~A (\kms)      & $      82.01       \pm  0.12      $ \\
Velocity amplitude for star~B (\kms)      & $     109.91       \pm  0.08      $ \\
Systemic velocity for star~A (\kms)       & $     -28.15       \pm  0.10      $ \\
Systemic velocity for star~B (\kms)       & $     -28.82       \pm  0.06      $ \\
{\it Derived parameters:} \\
Fractional radius of star~A               & $       0.13491    \pm  0.00052   $ \\       
Fractional radius of star~B               & $       0.0828     \pm  0.0015    $ \\       
Light ratio $\ell_{\rm B}/\ell_{\rm A}$   & $       0.255      \pm  0.010     $ \\
Orbital eccentricity                      & $       0.0036     \pm  0.0036    $ \\
Argument of periastron (\degr)            & $      89          \pm 90         $ \\[3pt]
\end{tabular}
\end{table}


Uncertainties were calculated using the Monte Carlo (MC) and residual-permutation (RP) simulations implemented in {\sc jktebop} \cite{Me08mn}, after the data uncertainties for each of the three dataset (TESS light curve and the RVs for each star) were scaled to give a reduced \chisq\ of $\chir = 1$. We find results in agreement with previous studies but with smaller errorbars relative to the most analogous work \cite{Catanzaro+24aa}. We could have decreased the errorbars further by fixing the LD coefficients and/or setting third light to zero, but such assumptions are not justified. The RP errorbars were gereally similar to but slightly larger than those from the MC simulations so were adopted. However, we retained the MC errorbars for the velocity amplitudes and systemic velocities because previous experience has shown that the RP errorbars are affected by small-number statistics \cite{Me21obs5}.

\begin{figure}[t] \centering \includegraphics[width=\textwidth]{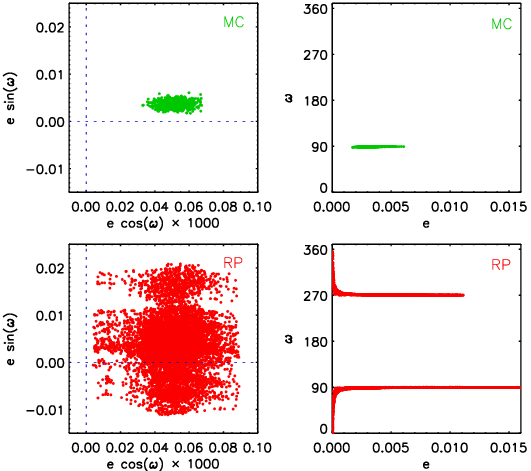} \\
\caption{\label{fig:ecc} Scatter plots of the MC (upper panels) and RP 
(lower panels) fits for the orbital shape parameters. Blue dotted lines 
indicate where $e\cos\omega$ and $e\sin\omega$ are zero.} \end{figure}

The amount of orbital eccentricity is questionable. Fig.~\ref{fig:ecc} shows that eccentricity is clearly detected using the MC error algorithm, but not for the RP algorithm; the difference is likely due to the noise characteristics of the data. $e\cos\omega$ is convincingly non-zero for both the MC and RP algorithms, indicating that the secondary eclipse is slightly later than phase 0.5. However, $e\sin\omega$ is consistent with zero given the available data. The situation is reflected in the sizes of the errorbars for the orbital shape parameters in Table~\ref{tab:jktebop}. A clearer understanding of these points would benefit from new data, but unfortunately there is currently no plan for TESS to revisit the field of \targ.

\section*{Physical properties and distance to \targ}

Although the {\sc jktebop} best fit includes calculated physical properties of the components of the system, we recalculated them using the {\sc jktabsdim} code \cite{Me++05aa} so we could use the RP errorbars when they were larger than the MC equivalents, excepting the spectroscopic properties listed above. We took all relevant properties from Table~\ref{tab:jktebop} and give the results in Table~\ref{tab:absdim}. 

Our measured properties of \targ\ agree well with those from previous works, with some caveats. The mass values are similar to those from refs.\ \citenum{KahramanAlicavus20raa} and \citenum{Catanzaro+24aa} but slightly smaller; roughly half the difference can be attributed to the larger orbital inclination found in this work, and half to differing velocity amplitudes. The radius values are also slightly smaller and significantly more precise. The comparison presented by ref.\ \citenum{Catanzaro+24aa} between their properties and those of ref.\ \citenum{KahramanAlicavus20raa} are erroneous because they compared their $M_{\rm A}$ and $M_{\rm B}$ with $M_{\rm A}\sin^3i$ and $M_{\rm B}\sin^3i$, which in the case of \targ\ is enough to cause significant disagreement, and because the comparison between the radii was based on a misreading or misinterpretation of the latter paper.

\begin{table} \centering
\caption{\em Physical properties of \targ\ defined using the nominal solar units 
given by IAU 2015 Resolution B3 (ref.~\citenum{Prsa+16aj}). \label{tab:absdim}}
\begin{tabular}{lr@{~$\pm$~}lr@{~$\pm$~}l}
{\em Parameter}        & \multicolumn{2}{c}{\em Star A} & \multicolumn{2}{c}{\em Star B}    \\[3pt]
Mass ratio   $M_{\rm B}/M_{\rm A}$          & \multicolumn{4}{c}{$0.7462 \pm 0.0012$}       \\
Semimajor axis of relative orbit (\Rsunnom) & \multicolumn{4}{c}{$14.441 \pm 0.011$}        \\
Mass (\Msunnom)                             &  1.6174 & 0.0034      &  1.2068 & 0.0036      \\
Radius (\Rsunnom)                           &  1.9482 & 0.0077      &  1.195  & 0.022       \\
Surface gravity ($\log$[cgs])               &  4.0676 & 0.0034      &  4.365  & 0.016       \\
Density ($\!\!$\rhosun)                     &  0.2187 & 0.0025      &  0.707  & 0.038       \\
Synchronous rotational velocity ($\!\!$\kms)& 26.06   & 0.10        & 15.99   & 0.29        \\
Effective temperature (K)                   &  7370   & 80          &  6410   & 80          \\
Luminosity $\log(L/\Lsunnom)$               &  1.004  & 0.019       &  0.337  & 0.027       \\
$M_{\rm bol}$ (mag)                         &  2.230  & 0.048       &  3.898  & 0.067       \\
Interstellar reddening \EBV\ (mag)          & \multicolumn{4}{c}{$0.04 \pm 0.02$}	        \\
Distance (pc)                               & \multicolumn{4}{c}{$125.0 \pm 1.5$}           \\[3pt]
\end{tabular}
\end{table}



We estimated the distance to \targ\ using the \Teff\ values from Carquillat et al.\ \cite{Carquillat+04mn}, $BV$ magnitudes from Tycho \cite{Hog+00aa} and $JHK_s$ magnitudes from 2MASS \cite{Cutri+03book} (see Table~\ref{tab:info}). The $JHK_s$ magnitudes were obtained at orbital phase 0.17 so are well away from eclipse. Application of the surface brightness calibrations from Kervella et al.\ \cite{Kervella+04aa} to all five passbands showed that an interstellar reddening of $\EBV = 0.04 \pm 0.02$~mag was needed to equalise the optical distance measurements with the infrared ones. Our final distance estimate is $125.0 \pm 1.5$~pc, which is in unimpeachable agreement with the $125.8 \pm 0.4$~pc from the \gaia\ DR3 parallax \cite{Gaia23aa}.

\section*{Comparison with theoretical models}

\begin{figure}[t] \centering \includegraphics[width=\textwidth]{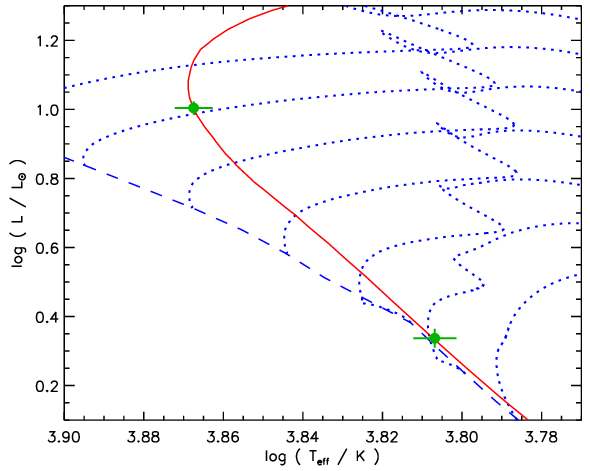} \\
\caption{\label{fig:hrd} Hertzsprung-Russell diagram for the components of \targ\ (filled 
green circles) and the predictions of the {\sc parsec} 1.2 models \cite{Bressan+12mn}. 
The dashed blue line shows the zero-age main sequence for a metallicity of $Z=0.014$. 
The dotted blue lines show evolutionary tracks for this metallicity and masses of 1.1\Msun\ 
to 1.7\Msun\ in steps of 0.1\Msun\ (from bottom-right to top-left). The solid red line shows 
an isochrone for this metallicity and an age of 1300~Myr.} \end{figure}

We compared the measured masses, radii, \Teff\ values and luminosities of the two stars to the predictions of the {\sc parsec} 1.2 theoretical stellar evolutionary models \cite{Bressan+12mn}. The stars are significantly different, resulting in a situation where the radii of the stars constrain the age well and their temperatures constrain the metallicity well. For all metallicities tested (specifically fractional metal abundances by mass, $Z$, between 0.010 and 0.030) the system age must be in the region of $1280 \pm 50$~Myr to match the radius of star~A for its mass. However, only the models for $Z=0.014$ can match the \Teff\ values, with a best age of 1300~Myr, indicating that the system has a mildly subsolar metallicity. An Hertzsprung-Russell diagram is shown in Fig.~\ref{fig:hrd}.

The chances of using \targ\ to assess the reliability of the theoretical models are limited by the chemical peculiarity of star~A, which makes its photospheric chemical abundances unrepresentative of its bulk metallicity. However, Catanzaro et al.\ \cite{Catanzaro+24aa} measured abundances for star~B which could be used to indicate the metallicity of both stars. These authors found abundances that are consistent with solar for 12 elements, and super-solar for five, which conflicts with our findings. The \Teff\ values presented by the three prior analyses of the system \cite{Carquillat+04mn,KahramanAlicavus20raa,Catanzaro+24aa} also differ by more than their uncertainties and none lead to a completely consistent agreement with theoretical models. We therefore advocate a new spectral analysis of \targ\ to confirm or resolve this discrepancy.

\section*{Summary and conclusions}

\targ\ is a dEB containing a slightly evolved Am star and a late-F star close to the zero-age main sequence. The 3.783-d orbit has a small but probably non-zero eccentricity, based on the orbital phase of secondary eclipse. Three studies of the system are available in the literature, all of which agree on the masses of the stars and the chemical peculiarity of the primary component, but none of which include precise measurements of the radii of the stars. We modelled the TESS sector 50 data to fill in this gap in knowledge of the system.

We determined the masses and radii to good precision -- 0.2\% and 0.3\% for mass and 0.4\% and 1.8\% for radius. The comparatively shallow partial eclipses prevent more precise radius measurements, and the difficulty in obtaining a reliable spectroscopic light ratio means it will be hard to improve on the current results. 

Our measurement of the distance of \targ\ is in excellent agreement with its \gaia\ DR3 parallax. The properties of the system match theoretical predictions for an age of 1300~Myr and a metallicity of $Z=0.014$; this sub-solar metallicity is discrepant with the measured photospheric chemical abundances of the secondary star. We searched for and found no evidence for pulsations in the system, in agreement with the suggestion that Am stars have a low fraction of pulsators \cite{Smalley+14aa,Smalley+17mn}.


\section*{Acknowledgements}

We thank the anonymous referee for a prompt report which led to much more discussion of the possible eccentricity of the system.
We thank Jerzy Kreiner and Waldemar Og{\l}oza for providing a list of times of minimum for \targ.
This paper includes data collected by the TESS\ mission and obtained from the MAST data archive at the Space Telescope Science Institute (STScI). Funding for the TESS\ mission is provided by the NASA's Science Mission Directorate. STScI is operated by the Association of Universities for Research in Astronomy, Inc., under NASA contract NAS 5–26555.


This work has made use of data from the European Space Agency (ESA) mission {\it Gaia}\footnote{\texttt{https://www.cosmos.esa.int/gaia}}, processed by the {\it Gaia} Data Processing and Analysis Consortium (DPAC\footnote{\texttt{https://www.cosmos.esa.int/web/gaia/dpac/consortium}}). Funding for the DPAC has been provided by national institutions, in particular the institutions participating in the {\it Gaia} Multilateral Agreement.
The following resources were used in the course of this work: the NASA Astrophysics Data System; the SIMBAD database operated at CDS, Strasbourg, France; and the ar$\chi$iv scientific paper preprint service operated by Cornell University.



\begin{thebibliography}{10}
\newcommand{\enquote}[1]{`(#1)'}

\bibitem{Andersen91aarv}
J.~{Andersen}, \textit{A\&ARv}, \textbf{3}, 91, 1991.

\bibitem{Torres++10aarv}
G.~{Torres}, J.~{Andersen} \& A.~{Gim{\'e}nez}, \textit{A\&ARv}, \textbf{18},
  67, 2010.

\bibitem{Me15aspc}
J.~{Southworth}, in \textit{Living Together: Planets, Host Stars and Binaries}
  (S.~M. {Rucinski}, G.~{Torres} \& M.~{Zejda}, eds.), 2015,
  \textit{Astronomical Society of the Pacific Conference Series}, vol. 496, p.
  321.

\bibitem{Me20obs}
J.~{Southworth}, \textit{The Observatory}, \textbf{140}, 247, 2020.

\bibitem{Me21univ}
J.~{Southworth}, \textit{Universe}, \textbf{7}, 369, 2021.

\bibitem{Abt61apjs}
H.~A. {Abt}, \textit{ApJS}, \textbf{6}, 37, 1961.

\bibitem{Abt65apjs}
H.~A. {Abt}, \textit{ApJS}, \textbf{11}, 429, 1965.

\bibitem{CarquillatPrieur07mn}
J.~{Carquillat} \& J.~{Prieur}, \textit{MNRAS}, \textbf{380}, 1064, 2007.

\bibitem{Hog+00aa}
E.~{H{\o}g} \textit{et~al.}, \textit{A\&A}, \textbf{355}, L27, 2000.

\bibitem{Cutri+03book}
R.~M. {Cutri} \textit{et~al.}, \textit{{2MASS All Sky Catalogue of Point
  Sources}} (The IRSA 2MASS All-Sky Point Source Catalogue, NASA/IPAC Infrared
  Science Archive, Caltech, US), 2003.

\bibitem{Gaia23aa}
{Gaia Collaboration}, \textit{A\&A}, \textbf{674}, A1, 2023.

\bibitem{CannonPickering20anhar}
A.~J. {Cannon} \& E.~C. {Pickering}, \textit{Annals of Harvard College
  Observatory}, \textbf{95}, 1, 1920.

\bibitem{ESA97}
ESA, \textit{ESA Special Publication}, \textbf{1200}, 1997.

\bibitem{Gaia21aa}
{Gaia Collaboration}, \textit{A\&A}, \textbf{649}, A1, 2021.

\bibitem{Stassun+19aj}
K.~G. {Stassun} \textit{et~al.}, \textit{AJ}, \textbf{158}, 138, 2019.

\bibitem{McGahee+20aj}
C.~{McGahee} \textit{et~al.}, \textit{AJ}, \textbf{160}, 52, 2020.

\bibitem{Carquillat+04mn}
J.~M. {Carquillat} \textit{et~al.}, \textit{MNRAS}, \textbf{352}, 708, 2004.

\bibitem{Kazarovets+99ibvs}
E.~V. {Kazarovets} \textit{et~al.}, \textit{IBVS}, \textbf{4659}, 1, 1999.

\bibitem{Bidelman88pasp}
W.~P. {Bidelman}, \textit{PASP}, \textbf{100}, 1084, 1988.

\bibitem{Grenier+99aas}
S.~{Grenier} \textit{et~al.}, \textit{A\&AS}, \textbf{137}, 451, 1999.

\bibitem{KahramanAlicavus20raa}
F.~{Kahraman Ali{\c{c}}avu{\c{s}}} \& F.~{Ali{\c{c}}avu{\c{s}}},
  \textit{Research in Astronomy and Astrophysics}, \textbf{20}, 150, 2020.

\bibitem{Catanzaro+24aa}
G.~{Catanzaro} \textit{et~al.}, \textit{A\&A}, \textbf{685}, A133, 2024.

\bibitem{Ricker+15jatis}
G.~R. {Ricker} \textit{et~al.}, \textit{Journal of Astronomical Telescopes,
  Instruments, and Systems}, \textbf{1}, 014003, 2015.

\bibitem{Jenkins+16spie}
J.~M. {Jenkins} \textit{et~al.}, in \textit{Proc.\ SPIE}, 2016, \textit{Society
  of Photo-Optical Instrumentation Engineers (SPIE) Conference Series}, vol.
  9913, p. 99133E.

\bibitem{Lightkurve18}
{Lightkurve Collaboration}, \enquote{{Lightkurve: Kepler and TESS time series
  analysis in Python}}, Astrophysics Source Code Library, 2018.

\bibitem{OteroDubovsky04ibvs}
S.~A. {Otero} \& P.~A. {Dubovsky}, \textit{IBVS}, \textbf{5557}, 1, 2004.

\bibitem{Porowski05ibvs}
C.~H. {Porowski}, \textit{IBVS}, \textbf{5606}, 1, 2005.

\bibitem{Brat+11oejv}
L.~{Brat} \textit{et~al.}, \textit{Open European Journal on Variable Stars},
  \textbf{137}, 1, 2011.

\bibitem{Zasche+11ibvs}
P.~{Zasche} \textit{et~al.}, \textit{IBVS}, \textbf{6007}, 1, 2011.

\bibitem{Pasche18oejv}
A.~{Pasche}, \textit{OEJV}, \textbf{191}, 1, 2018.

\bibitem{Pagel18ibvs}
L.~{Pagel}, \textit{IBVS}, \textbf{6244}, 1, 2018.

\bibitem{Pagel21bavj}
L.~{Pagel}, \textit{BAV Journal}, \textbf{052}, 1, 2021.

\bibitem{Paschke23bav}
A.~{Paschke}, \textit{BAV Journal}, \textbf{079}, 1, 2023.

\bibitem{Me++04mn2}
J.~{Southworth}, P.~F.~L. {Maxted} \& B.~{Smalley}, \textit{MNRAS},
  \textbf{351}, 1277, 2004.

\bibitem{Me13aa}
J.~{Southworth}, \textit{A\&A}, \textbf{557}, A119, 2013.

\bibitem{Hestroffer97aa}
D.~{Hestroffer}, \textit{A\&A}, \textbf{327}, 199, 1997.

\bibitem{Maxted18aa}
P.~F.~L. {Maxted}, \textit{A\&A}, \textbf{616}, A39, 2018.

\bibitem{Me23obs2}
J.~{Southworth}, \textit{The Observatory}, \textbf{143}, 71, 2023.

\bibitem{ClaretSouthworth22aa}
A.~{Claret} \& J.~{Southworth}, \textit{A\&A}, \textbf{664}, A128, 2022.

\bibitem{ClaretSouthworth23aa}
A.~{Claret} \& J.~{Southworth}, \textit{A\&A}, \textbf{674}, A63, 2023.

\bibitem{Kreiner++01book}
J.~M. {Kreiner}, C.-H. {Kim} \& I.-S. {Nha}, \textit{{An atlas of O-C diagrams
  of eclipsing binary stars}} (Wydawnictwo Naukowe AP, Krak\'ow, ul. Studencka
  5, Poland), 2001.

\bibitem{Me08mn}
J.~{Southworth}, \textit{MNRAS}, \textbf{386}, 1644, 2008.

\bibitem{Me21obs5}
J.~{Southworth}, \textit{The Observatory}, \textbf{141}, 234, 2021.

\bibitem{Me++05aa}
J.~{Southworth}, P.~F.~L. {Maxted} \& B.~{Smalley}, \textit{A\&A},
  \textbf{429}, 645, 2005.

\bibitem{Prsa+16aj}
A.~{Pr{\v s}a} \textit{et~al.}, \textit{AJ}, \textbf{152}, 41, 2016.

\bibitem{Kervella+04aa}
P.~{Kervella} \textit{et~al.}, \textit{A\&A}, \textbf{426}, 297, 2004.

\bibitem{Bressan+12mn}
A.~{Bressan} \textit{et~al.}, \textit{MNRAS}, \textbf{427}, 127, 2012.

\bibitem{Smalley+14aa}
B.~{Smalley} \textit{et~al.}, \textit{A\&A}, \textbf{564}, A69, 2014.

\bibitem{Smalley+17mn}
B.~{Smalley} \textit{et~al.}, \textit{MNRAS}, \textbf{465}, 2662, 2017.

\end{thebibliography}

\end{document}